\documentclass[a4paper,12pt,twoside]{article}

\usepackage{amssymb,latexsym,amsfonts,amsmath}
\usepackage{graphicx}
\usepackage[lined,algonl,boxed]{algorithm2e}

\newtheorem{theorem}{Theorem}

\newtheorem{definition}{Definition}

\newcommand{\R}{{\mathbb{R}}}

\title{
\vspace{-3cm}
{\LARGE  A Composition Theorem for\\ Bisimulation Functions} \\
\vspace{0.8cm}
{\Large Antoine Girard}\thanks{Laboratoire Jean Kuntzmann,Universit\'e Joseph Fourier, B.P. 53, 38041 Grenoble Cedex 9,
antoine.girard@imag.fr. This work has been supported by the Agence Nationale de la Recherche (VAL-AMS project - ANR-06-SETIN-018)}\\
\vspace{0.5cm}
{\large {\it Technical Note}}\\
\vspace{0cm}
\vspace{0.2cm}
{\large 2007}
}

\date{}

\begin{document}

\maketitle

\vspace{-2cm}

\begin{abstract}
The standard engineering approach to modelling of complex systems is highly compositional.
In order to be able to understand (or to control) the behavior of a complex dynamical systems,
it is often desirable, if not necessary, to view this system as an interconnection of smaller
interacting subsystems, each of these subsystems having its own functionalities.
In this paper, we propose a compositional approach to the computation of bisimulation functions for dynamical systems.
Bisimulation functions are quantitative generalizations of the classical
bisimulation relations. They have been shown useful for simulation-based verification or for the computation of approximate symbolic abstractions of dynamical systems. In this technical note, we present a constructive result for the composition of bisimulation functions.
For a complex dynamical system consisting of several interconnected subsystems,
it allows us to compute a bisimulation function from the knowledge of a bisimulation function for each of the subsystem.

\end{abstract}



\newpage
\section{Introduction}

The standard engineering approach to modelling of complex systems is highly compositional.
In order to be able to understand (or to control) the behavior of a complex dynamical systems,
it is often desirable, if not necessary, to view this system as an interconnection of smaller
interacting subsystems, each of these subsystems having its own functionalities. System
on chips, for instance, are often complex circuits that can be decomposed into smaller (and thus simpler)
circuits.

Albeit the simplification of the modelling process, a modular representation of complex systems
can greatly simplify the analysis process. In computer science, compositionallity and concurrency~\cite{Milner89}
have been a very active research field. In the system engineering science, a compositional approach is also
often used (see e.g.~\cite{Jiang96}). In this paper, we propose a compositional approach to the computation of bisimulation functions for dynamical systems.

Bisimulation functions have been introduced in~\cite{Girard2007}
as a quantitative generalization of the classical notion of
bisimulation relations that have been extensively and successfully
used in purely discrete systems analysis~\cite{Clarke99}. Bisimulation
functions measure how far two states of a system are from being
bisimilar, thus enabling the quantification of
the distance between trajectories originating from different states. 
Thus, these functions allow us to define a natural notion of neighborhood for trajectories of a system.
Recently, several promising papers have shown that bisimulation functions can be used for simulation-based verification~\cite{Girard06,Girard07a,Julius07,Lerda08} or for the computation of approximate symbolic abstractions of dynamical systems~\cite{Girard07,Pola07}. 

In this technical note, we present a constructive result for the composition of bisimulation functions.
For a complex dynamical system consisting of several interconnected subsystems,
it allows us to compute a bisimulation function from the knowledge of a bisimulation function for each of the subsystem.
Similar to Lyapunov functions for interconnected systems~\cite{Jiang96},
a small gain condition has to be fulfilled in order be able to compose bisimulation functions. The paper is organized
as follows. First, we present the notion of interconnection of subsystems useful for compositional modelling of dynamical
systems. Then, we introduce the notion of bisimulation function and develop a result on composition of bisimulation
functions.

\section{Compositional Modelling of Dynamical Systems}

Compositional modelling allows us to see a complex dynamical system $\Sigma$ 
as a set of several smaller subsystems $\Sigma_1,\dots,\Sigma_m$, interacting together.
This is a standard engineering approach and softwares such as Simulink or Scicos gained their
popularity from the possibility of modular representation of complex systems. 
In this section, we present the notion of interconnection of subsystems useful for compositional modelling of dynamical
systems.
In the following, we only define the interconnection of two subsystems; however, the extension to systems with 
more components is straightforward (see e.g.~\cite{Tazaki08}).

\begin{figure}[!h]
\begin{center}
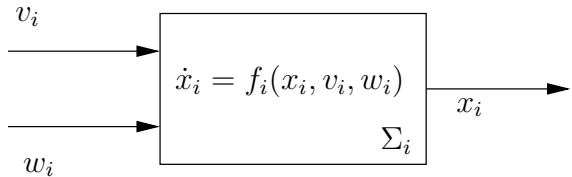
\end{center}
\caption{Subsystem $\Sigma_i$.}
\label{fig:ref1}
\end{figure}

Let us consider two dynamical systems, $\Sigma_1$ and $\Sigma_2$ of the following form:
$$
\Sigma_i: \dot x_i(t) =f_i(x_i(t),u_i(t)), \; i=1,2.
$$
where $x_i(t)\in \R^{n_i}$ and $u_i(t)\in \R^{m_i}$ denote the state and input variables of $\Sigma_i$.
The input vector is of the form $u_i(t)=[v_i(t),w_i(t)]$, where $v_i(t)\in \R^{p_i}$ denotes the inputs used for the interconnection of $\Sigma_1$ and $\Sigma_2$ and $w_i(t)\in \R^{q_i}$ denotes the external inputs (see Figure~\ref{fig:ref1}). 

\begin{figure}[!h]
\begin{center}
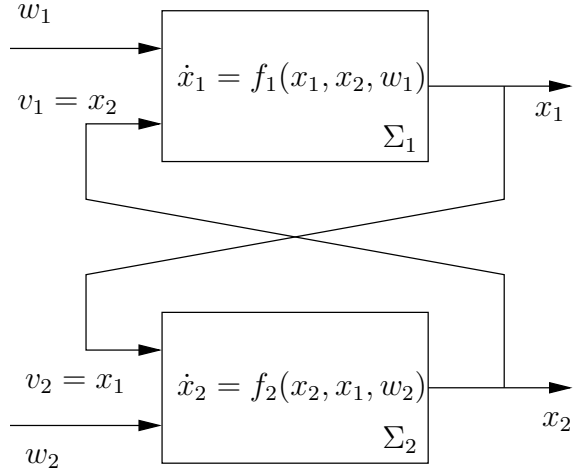
\end{center}
\caption{The composition of $\Sigma_1$ and $\Sigma_2$.}
\label{fig:ref2}
\end{figure}

The interconnection of $\Sigma_1$ and $\Sigma_2$ is achieved by feeding the system inputs $v_1(t)$ and $v_2(t)$ with the state variables $x_2(t)$ and $x_1(t)$ (see Figure~\ref{fig:ref2}). We therefore assume that $p_1=n_2$ and $p_2=n_1$. 
Then, the interconnection of $\Sigma_1$ and $\Sigma_2$ is formally defined as follows:
\begin{definition} The interconnection of $\Sigma_1$ and $\Sigma_2$ is the dynamical
system $\Sigma$ given by the differential equation
$$
\Sigma : 
\left\{
\begin{array}{lll}
\dot x_1(t) &=&f_1(x_1(t),x_2(t),w_1(t)),\\
\dot x_2(t) &=&f_2(x_2(t),x_1(t),w_2(t))
\end{array}
\right.
$$
\end{definition}
The state of $\Sigma$ is $x(t) = [x_1(t),x_2(t)]\in \R^n$ with $n=n_1+n_2$ and the input of
$\Sigma$ is $u(t)=[w_1(t),w_2(t)]\in \R^m$ with $m=q_1+q_2$. Then, the system $\Sigma$ can be written under the form
$$
\Sigma: \dot x(t) =f(x(t),u(t)),
$$
that is similar to $\Sigma_1$ and $\Sigma_2$. Then, this means that $\Sigma$ can be composed with another system, enabling the hierarchical modelling of dynamical systems.

\section{Composition of Bisimulation Functions}

We first present the notion of bisimulation function, then we will give a result on composition of bisimulation
functions.

\subsection{Bisimulation functions}

Let us consider a dynamical system of the form 
$$
\Sigma: \dot x(t) =f(x(t),u(t))
$$
where $x(t)\in \R^n$ and $u(t)\in \R^m$.
Bisimulation functions have been introduced in~\cite{Girard2007}
as a quantitative generalization of the classical notion of
bisimulation relations that have been extensively and successfully
used in purely discrete systems analysis~\cite{Clarke99}. Bisimulation
functions measure how far two states of a system are from being
bisimilar, thus enabling the quantification of
the distance between trajectories originating from different states. 
Thus, these functions allow us to define a natural notion of neighborhood for trajectories of a system.
The following definition slightly differs from the original definition in~\cite{Girard2007}.
It is the continuous time version of the definition given in~\cite{Girard07a} which makes it suitable for simulation-based 
verification. 
\begin{definition} A smooth function $V:\R^n \times \R^n \rightarrow \R^+$ is a bisimulation function for $\Sigma$ 
if 
\begin{equation}
\label{eq:bis1}
\|x-x'\| \le V(x,x')
\end{equation}
and there exists $\lambda >0$, $\gamma \ge 0$ such that for all $x\in \R^n$, $x'\in \R^n$, $\forall u\in \R^m, u'\in \R^m$,
\begin{equation}
\label{eq:bis2}
\frac{\partial V}{\partial x} f(x,u) +\frac{\partial V}{\partial x'} f(x',u') \le -\lambda V(x,x') + \gamma \|u-u'\|.
\end{equation}
\end{definition}

Bisimulation functions have the following property which makes them suitable tools for simulation-based verification~\cite{Girard06,Girard07a,Julius07,Lerda08}
or for the computation of approximate symbolic abstractions of dynamical systems~\cite{Girard07,Pola07}.

\begin{theorem} Let us consider $x(t)$ and $x'(t)$ be the trajectories of $\Sigma$ given by
$$
\dot x(t)= f(x(t),u(t)) \text{ and } \dot x'(t)= f(x'(t),u'(t)).
$$
Then, we have for all $t\ge 0$
$$
\|x(t)-x'(t)\| \le V(x(t),x'(t)) \le e^{-\lambda t} V(x(0),x'(0)) + \frac{\gamma}{\lambda} \|u-u'\|_\infty
$$
where $\|u-u'\|_\infty = \sup_{t \ge 0} \|u(t)-u'(t)\|$.
\end{theorem}

{\it Proof}~: From equation (\ref{eq:bis1}), we have the first inequality. From equation (\ref{eq:bis2}), we have
$$
\frac{dV(x(t),x'(t))}{dt}  \le -\lambda V(x(t),x'(t)) + \gamma \|u(t)-u'(t)\| \le -\lambda V(x(t),x'(t)) + \gamma \|u-u'\|_\infty
$$
Let $\eta(t)=e^{-\lambda t} V(x(0),x'(0)) + \frac{\gamma}{\lambda}  \|u-u'\|_\infty$, it is a solution of the differential equation
$$
\dot \eta(t) = -\lambda \eta(t) + \gamma \|u-u'\|_\infty.
$$
Moreover, $V(x(0),x'(0)) \le \eta(0)$; then, from the funnel theorem (see e.g.~\cite{Hubbard95}), it follows that for all $t\ge 0$,
$V(x(t),x'(t)) \le \eta(t)$.
$\blacksquare$ 

The practical computation of bisimulation functions is out of the scope of this technical note. However, we refer the interested reader to~\cite{Girard05,Girard07b} for computational methods applying to linear and nonlinear dynamical systems. 

Let us remark that the previous theorem clearly shows the existing relation
between the notion of bisimulation function and the notion of incremental input-to-state stability~\cite{angeli2002}
(close initial states and close inputs lead to close trajectories of $\Sigma$).
This connection was already pointed out in the work~\cite{Pola07} where incremental input-to-state stability
was shown sufficient for the existence of approximately bisimilar symbolic abstractions of a dynamical system.

\subsection{A Composition Result for Bisimulation Functions}

We now consider the problem of composing bisimulation functions. For complex systems that consists of several interconnected subsystems, it is interesting to develop compositional analysis methods. Let us assume that we are given a bisimulation function for each subsystem, then the question is whether it is possible or not to compose these functions
to design a bisimulation function for the global system. The following result shows that the composition is possible under a small gain condition. It has similarities with~\cite{Jiang96} where a compositional result for ISS-Lyapunov functions
is developped.

\begin{theorem} Let $\Sigma_1$ and $\Sigma_2$ be dynamical systems and let $\Sigma$ be the interconnection of $\Sigma_1$ and $\Sigma_2$ as defined in Definition 1. Let $V_1$ and $V_2$ be simulation functions for $\Sigma_1$ and $\Sigma_2$,
we denote by $\lambda_1$ and $\gamma_1$ (respectively $\lambda_2$ and $\gamma_2$) the real numbers such that equation (\ref{eq:bis2}) holds for $V_1$ (respectively $V_2$). 
Then, under the small gain condition $\frac{\gamma_1 \gamma_2}{\lambda_1 \lambda_2} < 1$, there exists $V$ a bisimulation
function for $\Sigma$ of the form:
\begin{equation}
\label{eq:V}
V(x,x')=\alpha_1 V_1(x_1,x_1') + \alpha_2 V_2(x_2,x_2') \text{ where } x=[x_1,x_2],\; x'=[x_1',x_2'].
\end{equation}
The couple $(\alpha_1,\alpha_2)$ can be chosen as follows
\begin{equation}
\label{eq:alpha}
\left\{
\begin{array}{llll}
\frac{\gamma_2}{\lambda_1}<\alpha_1<\frac{\lambda_2}{\gamma_1}& \text{ and } &
\alpha_2=1 & \text{ if } \lambda_1 \le \gamma_2 \\
\alpha_1=1 & \text{ and } & \frac{\gamma_1}{\lambda_2} < \alpha_2 < 
\frac{\lambda_1}{\gamma_2} & \text{ if } \lambda_2 \le \gamma_1 \\
\alpha_1=1 & \text{ and } & \alpha_2=1 & \text{in the other cases}.
\end{array}
\right.
\end{equation}

\end{theorem}

\begin{figure}[!h]
\begin{center}
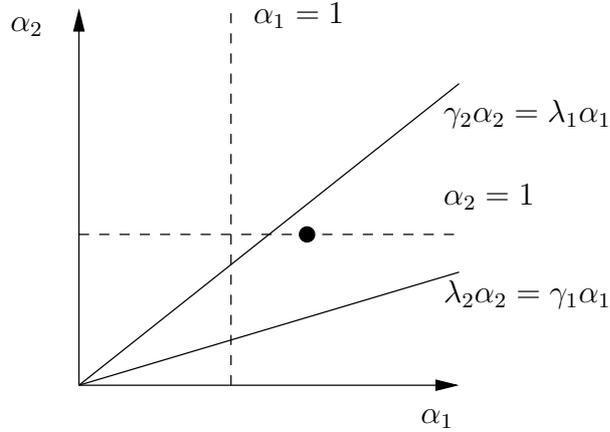
\end{center}
\caption{Configuration 1: $\lambda_1 \le \gamma_2$.}
\label{fig:case1}
\end{figure}

\begin{figure}[!h]
\begin{center}
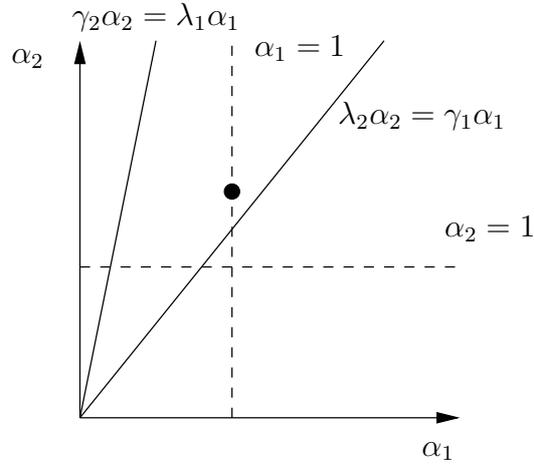
\end{center}
\caption{Configuration 2: $\lambda_2 \le \gamma_1$.}
\label{fig:case2}
\end{figure}

\begin{figure}[!h]
\begin{center}
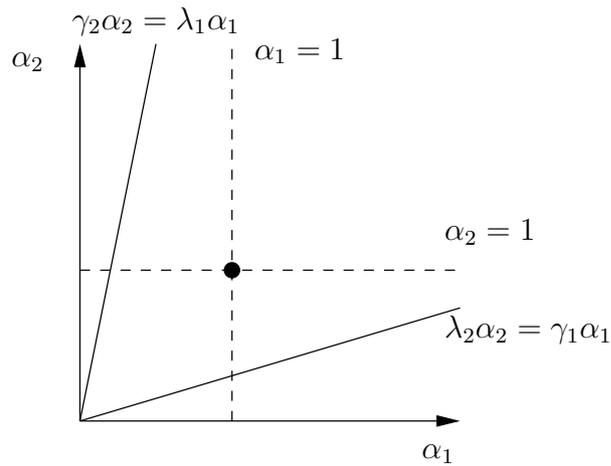
\end{center}
\caption{Configuration 3: other cases.}
\label{fig:case3}
\end{figure}

{\it Proof}: Let $V$ be a function of the form (\ref{eq:V}), we look for conditions on $\alpha_1$
and $\alpha_2$ ensuring that $V$ is a bisimulation function for $\Sigma$. First, let us remark that
if $\alpha_1 \ge 1$ and $\alpha_2 \ge 1$ then,
$$
V(x,x') \ge V_1(x_1,x_1') + V_2(x_2,x_2') \ge \|x_1-x_1'\|+\|x_2-x_2'\|
$$
because $V_1$ and $V_2$ satisfy equation (\ref{eq:bis1}). Then, by remarking that
$$
\|x-x'\|=\sqrt{\|x_1-x_1'\|^2+\|x_2-x_2'\|^2} \le  \|x_1-x_1'\|+\|x_2-x_2'\|,
$$
it follows that $V$ satisfies equation  (\ref{eq:bis1}) as well. 
Let $u=[w_1,w_2]$, $u'=[w_1',w_2']$ be inputs of $\Sigma$. Then, we have
\begin{eqnarray*}
\frac{\partial V}{\partial x} f(x,u) +\frac{\partial V}{\partial x'} f(x',u') 
&=& \alpha_1\frac{\partial V_1}{\partial x} f_1(x_1,x_2,w_1) +\alpha_2\frac{\partial V_2}{\partial x} f_2(x_2,x_1,w_2)\\
& & +\alpha_1\frac{\partial V_1}{\partial x'} f_1(x'_1,x'_2,w'_1)+\alpha_2 \frac{\partial V_2}{\partial x'} f_2(x'_2,x'_1,w'_2) \\
&\le& \alpha_1 \left(-\lambda_1 V_1(x_1,x_1') +\gamma_1 \|[x_2,w_1]-[x'_2,w'_1]\|\right)\\
& & 
+\alpha_2\left(-\lambda_2 V_2(x_2,x_2') + \alpha_2\gamma_2 \|[x_1,w_2]-[x'_1,w'_2]\|\right)
\end{eqnarray*}
because $V_1$ and $V_2$ satisfy equation (\ref{eq:bis2}). Further, we have
$$
\|[x_2,w_1]-[x'_2,w'_1]\|=\sqrt{\|x_2-x_2'\|^2+\|w_1-w_1'\|^2} \le  \|x_2-x_2'\|+\|w_1-w_1'\| 
$$
and
$$
\|[x_1,w_2]-[x'_1,w'_2]\|=\sqrt{\|x_1-x_1'\|^2+\|w_2-w_2'\|^2} \le  \|x_1-x_1'\|+\|w_2-w_2'\|. 
$$
Therefore,
\begin{eqnarray*}
\frac{\partial V}{\partial x} f(x,u) +\frac{\partial V}{\partial x'} f(x',u') &\le &
\alpha_1\left(-\lambda_1 V_1(x_1,x_1') + \gamma_1 \|x_2-x_2'\| + \gamma_1 \|w_1-w_1'\|\right) \\
& & 
+\alpha_2\left(-\lambda_2 V_2(x_2,x_2') + \gamma_2 \|x_1-x'_1\|+ \gamma_2 \|w_2-w'_2\|\right).
\end{eqnarray*}
Then, since $V_1$ and $V_2$ satisfy equation (\ref{eq:bis1}), it follows that
\begin{eqnarray*}
\frac{\partial V}{\partial x} f(x,u) +\frac{\partial V}{\partial x'} f(x',u') &\le &
\alpha_1\left(-\lambda_1 V_1(x_1,x_1') + \gamma_1 V_2(x_2,x_2') + \gamma_1 \|w_1-w_1'\|\right) \\
& & 
+\alpha_2\left(-\lambda_2 V_2(x_2,x_2') + \gamma_2 V_1(x_1,x_1')+ \gamma_2 \|w_2-w'_2\|\right) \\
&\le &
-(\alpha_1\lambda_1 - \alpha_2\gamma_2 )V_1(x_1,x_1') + \alpha_1\gamma_1 \|w_1-w_1'\| \\
& & 
-(\alpha_2\lambda_2-\alpha_1\gamma_1) V_2(x_2,x_2') + \alpha_2\gamma_2 \|w_2-w'_2\|.
\end{eqnarray*}
Let us assume that $\alpha_1\lambda_1 - \alpha_2\gamma_2>0$ and $\alpha_2\lambda_2-\alpha_1\gamma_1>0$, then let us define
$$
\lambda=\min\left( \frac{\alpha_1\lambda_1 - \alpha_2\gamma_2}{\alpha_1} , \frac{\alpha_2\lambda_2-\alpha_1\gamma_1}{\alpha_2}
\right) \text{ and } \gamma =\alpha_1 \gamma_1+ \alpha_2 \gamma_2.
$$
By remarking that $\|w_1-w_1'\| \le \|u-u'\|$ and $\|w_2-w_2'\| \le \|u-u'\|$ it follows that
$$
\frac{\partial V}{\partial x} f(x,u) +\frac{\partial V}{\partial x'} f(x',u') \le -\lambda V(x,x') +\gamma \|u-u'\|.
$$
Therefore, we proved that if $\alpha_1 \ge 1$, $\alpha_2 \ge 1$, $\alpha_1\lambda_1 - \alpha_2\gamma_2>0$ and $\alpha_2\lambda_2-\alpha_1\gamma_1>0$, then $V$ is a bisimulation function for $\Sigma$. 
Let us show that a necessary and sufficient condition for the existence of a couple $(\alpha_1,\alpha_2)$ satisfying
these four inequalities is
$\frac{\gamma_1 \gamma_2}{\lambda_1 \lambda_2} < 1$. Let the inequalities hold, then particularly,
$$
\alpha_1\lambda_1\lambda_1> \alpha_2\gamma_2\lambda_2> \alpha_1 \gamma_1 \gamma_2.
$$
It follows that necessarily $\frac{\gamma_1 \gamma_2}{\lambda_1 \lambda_2} < 1$.
Conversely, if $\frac{\gamma_1 \gamma_2}{\lambda_1 \lambda_2} < 1$, there are only three possible configurations
shown on Figures~\ref{fig:case1},~\ref{fig:case2} and~\ref{fig:case3}. Then, by choosing $\alpha_1$ and $\alpha_2$
as in equation (\ref{eq:alpha}), the four inequalities hold.~$\blacksquare$

This theorem provides us with a method to compute compositionally bisimulation functions
for composite systems. Note that it is subject to a small gain condition that is $\frac{\gamma_1 \gamma_2}{\lambda_1 \lambda_2} < 1$. Let us remark that the choice of the couple $(\alpha_1,\alpha_2)$ given in equation (\ref{eq:alpha})
is only one possible choice among many others satisfying the inequalities $\alpha_1 \ge 1$, $\alpha_2 \ge 1$, $\alpha_1\lambda_1 - \alpha_2\gamma_2>0$ and $\alpha_2\lambda_2-\alpha_1\gamma_1>0$. Another choice might be more suitable depending on the considered application.

\section{Conclusions}
In this technical note, we presented a constructive result for the composition of bisimulation functions.
For a complex dynamical system consisting of several interconnected subsystems,
it allows us to compute a bisimulation function from the knowledge of a bisimulation function for each of the subsystem.
Similar to Lyapunov functions for interconnected systems,
a small gain condition has to be fulfilled in order be able to compose bisimulation functions.

In the context of the VAL-AMS project, this result shall be useful for the computation of bisimulation functions for large scale analog circuits which can be seen as the interconnection of smaller circuits.
The knowledge of a bisimulation function is required for simulation-based verification~\cite{Girard06,Girard07a,Julius07,Lerda08}
or for the computation of approximate symbolic abstractions of dynamical systems~\cite{Girard07,Pola07}.

\bibliography{rapport}
\bibliographystyle{plain}

\end{document}